# Towards Sustainable Energy-Efficient Data Centers in Africa

David Ojika
*Flapmax*
Austin, Texas
dave@flapmax.com

Jayson Strayer
*Intel*
Hillsboro, Oregon
jayson.strayer@intel.com

Gaurav Kaul
*HPE*
London, UK
gaurav.kaul@hpe.com

*Abstract*—Developing nations are particularly susceptible to the adverse effects of carbon emissions. By 2040, 14% of world's emissions will come from data centers. This paper presents early findings in the use AI and digital twins to model and optimize data center operations.

## I. Introduction

Global warming is one of the most contentious international debates that have been heard in both private and public arenas. This is as a result of the different effects global warming has on livelihood of all socioeconomic levels. Developing countries are particularly susceptible to human health crises, rising sea levels, changing weather patterns, risks of flooding or drought as well as a threats to natural biodiversity [1]. All these concerns have driven more awareness in the technology industry, especially in cloud computing and cryptocurrency mining, which have seen significant increase in usage, resulting in an increase in the amount of cooling required by these expansive server farms. (40% of the total energy goes to cooling IT equipment [7]). Cooling of such facilities requires significant energy which in effect leads to increased carbon emissions. Thus, in line with the White House 2030 Greenhouse Gas Pollution Reduction Target [2], the overarching vision of this research is to both quantify and aim to achieve significant $CO_2$ emission reduction in data centers, over the current decade.

Towards realizing this vision, we propose a systemic approach that will work to reduce carbon emissions of data centers by modeling and simulating energy usage and predicting equipment failure. This is achieved using low-cost sensors coupled with Internet of Things (IoT) gateway devices and advanced Artificial Intelligence (AI) processing capabilities. Through these technologies, the proposed system aims to provide several analytical predictions that optimize energy demand (to avoid wastage of resources), reduce overall costs, enhance operational efficiency (of the workers because they now have better data-driven tools, better insights, hence better decision-making) among others. We focus on the problem of unpredictable data-center building maintenance and anomalies and propose the use of AI and digital twins to simulate different scenarios to enable the management staff to optimize the use of such facilities. In the rest of this paper, we will discuss our methodology for this design, present our findings, and conclude with an outlook on the past and future carbon footprint in Africa, the world's fastest growing economy.

## II. Methodology

Our methodology involves the design of a digital twin model that models a data center through data collected from sensors and the data center's building design and use that data to predict the likelihood of an anomaly as well as perform what-if simulations. This design combines two emerging technologies: Graph Neural Networks (GNN) and Deep Reinforcement Learning (DLR). The GNN models the building structure, presenting a graphical representation of floors, IoT devices, and sensors connected to IT equipment and HVAC systems, whereas the DRL continuously trains the GNN model. This training is important to reflect ambient changes (e.g., temperature, humidity, etc.) that may have occurred after the graph is generated. Additionally, the methodology provides a user interface where a user may interact with the GNN/DRL combination (i.e., the digital twin model), providing inputs to the model and receiving feedback and prediction scores. For these scores, one may consider a simple binary classification (i.e., stable or anomaly), percent score (e.g., 60% chance of anomaly), or individuals scores for specific failures (e.g., 30% chance of air quality issue).

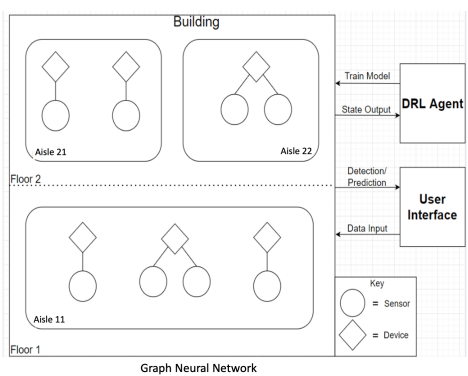

Fig 1a. A simplified digital twin and graphical representation of a data center (Fig 1b) showing the DRL agent and UI. Sensors may connect to OCP's *DeviceManager* which in turn interfaces with the digital twin.

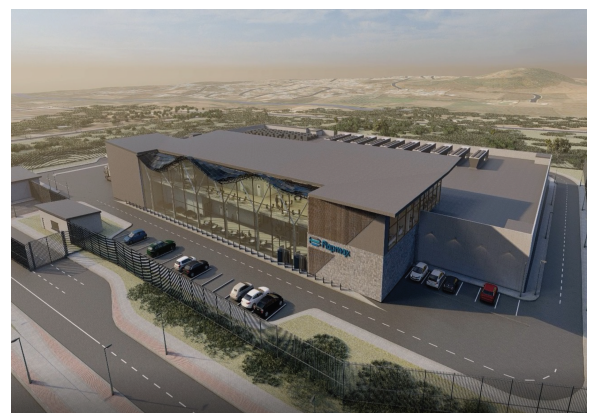

Fig 1b. A 3D model of a "smart data center" under study. The data center design is inspired by Open Compute Project (OCP).

Developing a GNN and DRL model are both difficult tasks and having them interact with one another is an even more difficult task. To reduce the risk of any errors later, we developed prototypes for both the GNN and DRL model and tested them for correctness and functionality. As illustrated in Fig 1, our current design contains a basic GNN model and DRL agent that together model and simulate a 2-floor data center building through its various components (e.g., HVAC, age, size, activities) to predict potential faults and anomalies within the building. The DRL model periodically trains the GNN model as new inputs and results come in.

### III. EXPERIMENT RESULTS

Our experiment was conducted in two parts: testing and evaluation of the GNN model, and performance tuning of the DRL agent. For the GNN experiment, using proprietary data sets we trained the GNN model up to 100 epochs under cross-entropy loss and obtained very promising results (training accuracy: 0.911, testing accuracy: 0.907, training ROC_AUC: 0.724, and testing ROC_AUC: 0.728). For the DRL experiment, we evaluated various search-optimization algorithms, including (Grid search, Ax, SkOpt, and HyperOpt) across different hyperparameters and state saving variables. Our findings indicate that while SkOpt slightly outperforms HyperOpt, HyperOpt has higher potential because of state saving. State saving also offers both short term improvements and promises better gains the more the model runs.

Beyond these data science and deep learning concepts, correct performance of the GNN/DRL-based digital twin model relies on several other factors (e.g., consistency in handling various data type as well as computing, storage, and networking capabilities). Because our project is still in early development and the findings that we present are preliminary results, it is important to tweak the model further in different ways and examine how the model's performance is impacted under specific geographies.

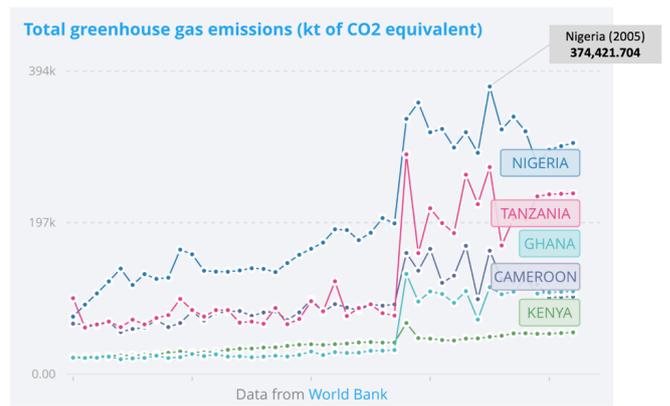

Fig 2. Carbon footprint in five strategic countries in Africa.

### IV. AN OUTLOOK ON AFRICA

Africa as a continent is mainly composed of developing nations. The different countries are therefore developing their carbon footprint portfolio as much as every other economic aspect of their sovereign regions. With a benchmark of 2005 (Fig 2), Africa is yet to reach alarming levels of carbon footprint. As illustrated, different regions can have different levels of carbon footprint hence the need to be strategic about the placement of a data center in those regions. After choosing a location it is important that systems are in place to continue to monitor carbon emissions, optimize the use of IT/cooling resources through automation, and accommodate future upgrades to realize the overall goal of reducing global warming. This may only be achieved through improved connectivity and analytics that enable data sharing more efficiently hence reducing the need to compute the same in different regions.

### ACKNOWLEDGMENT

The authors acknowledge the contributions of Ahmed Shafaat Ahsen, James Lu, John Nguyen, and Tyler Black from the College of Natural Sciences at the University of Texas at Austin and thank both Intel and HPE for their hardware donations.

Table 1. Performance tuning of the DRL agent.

| Experiment# | Algorithm | Hyperparameters | State Saving | Loss |
|---|---|---|---|---|
| 1 | Grid | Learning Rate | No | 0.0249374 |
| 2 | Grid | Momentum | No | 0.03144178 |
| 3 | Grid | LR + M | No | 0.02139868 |
| 4 | Ax | Learning Rate | No | 0.0018266 |
| 5 | Ax | Momentum | No | 0.006096561 |
| 6 | Ax | LR + M | No | 0.0013472 |
| 7 | Sk0pt | Learning Rate | No | 0.00075684 |
| 8 | Sk0pt | Momentum | No | 0.00097632 |
| 9 | Sk0pt | LR + M | No | 0.00011701 |
| 10 | HyperOpt | Learning Rate | No | 0.000767532 |
| 11 | HyperOpt | Momentum | No | 0.00264834 |
| 12 | HyperOpt | LR + M | No | 0.000135889 |
| 13 | HyperOpt | Learning Rate | Yes | 0.000254122 |
| 14 | HyperOpt | Momentum | Yes | 0.000705588 |
| 15 | HyperOpt | LR + M | Yes | 0.000120037 |